\documentclass[amsmath,amssymb,showpacs,draft,preprint,floatfix]{revtex4}
\usepackage{graphicx}
\usepackage{latexsym}
\usepackage{amsmath}
\usepackage{amsfonts}
\usepackage{amssymb}
\begin{document}
\title{Quantum derivation of Manley Rowe type relations}

\author{M. Fern\'andez Guasti}

\affiliation{Departamento de F\'{\i}sica,
Universidad Aut\'onoma Metropolitana - Iztapalapa, 09340 M\'exico
D.F., Apartado postal 55-534, Mexico. }

\author{H. Moya-Cessa}

\affiliation{Instituto Nacional de Astrf\'{\i}sica Optica y
Electr\'onica, Coordinaci\'on de Optica, Apdo. Postal 51 y 216,
72000 Puebla, Pue., Mexico\\ E-mail: hmmc@inaoep.mx}

\begin{abstract}
 The Ermakov Lewis quantum invariant for the time
dependent harmonic oscillator is expressed in terms of number and
phase operators. The identification of these variables is made in
accordance with the correspondence principle and the amplitude and
phase representation of the classical orthogonal functions
invariant. The relationship between the number and phase operators
is established through this invariant as the system evolves from
one frequency to another. In the specific case where the
excitations represent the photon number, these relations are
equivalent to the power density transport equations derived in
nonlinear optical processes.
\end{abstract}
\maketitle
\section{Introduction}

Different exact invariants have been used either to solve the one
dimensional time dependent Shr\"{o}dinger equation  \cite{Ray} or to
obtain the general differential equation solution from a
particular one  \cite{Lutzky}. The Ermakov-Lewis invariant and the
orthogonal functions invariants are two schemes that have been
used to solve the quantum mechanical time dependent harmonic
oscillator equation in one dimension (QM-TDHO)  \cite{gmc}. These
invariants are usually expressed in terms of coordinate and
momentum operators. Nonetheless, an amplitude and phase operators
representation of these invariants is also possible. However, such
a representation is not unique and furthermore, it is not straight
forward to associate a number operator representing the number of
excitations to these variables  \cite{fermoy pra}.

On the other hand, despite the mathematical success of these
formalisms there is scarcely any literature regarding the physical
interpretation of these quantitites. Two exceptions, to the best
of our knowledge, are the proposal made by Eliezer  \cite{Eliezer}
regarding the Ermakov-Lewis classical invariant and the
interpretation of the orthogonal functions invariant given by
us  \cite{fer2}.

The conservation of electromagnetic field power density in
non-linear processess is described by the Manley-Rowe
relations  \cite{Manley}. These equations have been obtained in
semiclassical theory using the nonlinear wave equation derived
from Maxwell's equations as a starting point. The participating
waves are considered harmonic in time, colinear and the slowly
varying spatial envelope approximation is imposed in order to
obtain a set of coupled first order linear differential
equations  \cite{Yariv}. The nonlinear susceptibility tensor for the
specific wave mixing process is then evaluated in a lossless
medium and its symmetries are established through the Kleinmann's
conditions  \cite{Shen}. The relations thus obtained are interpreted
in terms of the conservation of the photon flux as the fields
propagate in space.

In this presentation, we first recreate the procedure used to
obtain the amplitude and phase representation in the classical and
quantum cases. To this end, the orthogonal functions and the
Ermakov-Lewis invariants are described trying to stress the
physical meaning of the variables involved. The phase and number
operators are then associated according with the role that the
variables play. The Ermakov-Lewis quantum invariant is shown to
represent the energy conservation of the closed oscillator
ensemble.

The formalism is applied to a photon field in an analogous fashion
as the black body radiation problem was tackled by
Planck  \cite{Planck}. Namely, the field properties are obtained
from the oscillator ensemble with which the radiation interacts. A
quantum derivation of the conservation equations is then obtained
for arbitrary non-linear frequency conversion processes.

\section{Classical and Quantum invariants}

\subsection{Orthogonal functions invariant}

Consider the time dependent Schr\"{o}dinger equation with
$\hbar=1$
\begin{equation}
i\frac{\partial|\psi(t)\rangle}{\partial t}=\hat{H}|\psi(t)\rangle,\label{tih}%
\end{equation}
for a time dependent harmonic oscillator Hamiltonian
$\hat{H}(t)=1/2\left( \hat{p}^{2}+\Omega^{2}(t)\hat{q}^{2}\right)
$. The quantum orthogonal
functions invariants of this system are%
\begin{equation}
\hat{G}_{1}=u_{1}\hat{p}-\dot{u}_{1}\hat{q},\qquad\hat{G}_{2}=-u_{2}\hat
{p}+\dot{u}_{2}\hat{q}.\label{gs}%
\end{equation}
These invariants obey the commutation relation%
\begin{equation}
\lbrack\hat{G}_{1},\hat{G}_{2}]=-iG;
\end{equation}
$G$ is a constant given by the classical orthogonal functions
invariant  \cite{fer1}
\begin{equation}
G=u_{1}\dot{u}_{2}-u_{2}\dot{u}_{1},\label{inv G classical}%
\end{equation}
where $u_{1}$ and $u_{2}$ are real linearly independent solutions
of the TDHO equation $\ddot{u}+\Omega^{2}(t)u=0$. The amplitude
$\rho$ and phase $s_{\rho }$ representation of this
\textit{c}-number is straight forward from the
substitution of $u_{1}=-\rho\sin s_{\rho}$, $u_{2}=\rho\cos s_{\rho}$%
\begin{equation}
G=\rho^{2}\dot{s}_{\rho}.
\end{equation}
Let the constant $G$ at an initial time be given by%
\begin{equation}
G=\rho_{0}^{2}\omega_{0},
\end{equation}
where the derivative of the phase is defined as the frequency
$\omega\left( t\right)  \equiv\dot{s}_{\rho}$. The squared
amplitude times the frequency at any subsequent time obey the
relationship
\begin{equation}
\rho^{2}\left(  t\right)  \omega\left(  t\right)
=\rho_{0}^{2}\omega
_{0},\label{rho sq omeg =1}%
\end{equation}
$\rho$ satisfies the differential equation $\ddot{\rho}+\Omega^{2}%
(t)\rho=1/\rho^{3}$.

\subsection{Ermakov Lewis Invariant}

The Ermakov Lewis invariant may be written as  \cite{Lewis}%

\begin{equation}
\hat{I}=\frac{1}{2}\left(  \hat{G}_{1}^{2}+\hat{G}_{2}^{2}\right)
\label{I gs}%
\end{equation}%
\begin{equation}
=\frac{1}{2}\left[  \left(  \frac{G\hat{q}}{\rho}\right)
^{2}+(\rho\hat
{p}-\dot{\rho}\hat{q})^{2}\right]  ,\label{I lewis}%
\end{equation}
where the amplitude function is $\rho=\sqrt{u_{1}^{2}+u_{2}^{2}}$.
The classical Ermakov Lewis constant of motion in the amplitude
and phase representation follows from the substitution
$\hat{q}\rightarrow\rho\cos s_{\rho}$, $\hat{p}\rightarrow
d\hat{q}/dt$ :
\begin{equation}
I=\frac{1}{2}\rho^{4}\dot{s}_{\rho}^{2}.\label{lewis inv amph}%
\end{equation}
This expression, according to Eliezer and Gray  \cite{Eliezer}\ may
be interpreted as the square of an angular momentum through the
introduction of an auxiliary imaginary axis perpendicular to the
actual direction of the oscillator motion.

\section{Creation and annihilation operators}

An operator that can be written as the sum of two squares may be
expressed in terms of two adjoint complex quantities. The
invariant operator $\hat{I}$ may be written as the sum of two
squares in two different ways, namely (\ref{I gs}) and (\ref{I
lewis}). The former expression leads to annihilation and creation
operators of the form
\begin{equation}
\hat{A}=\frac{1}{\sqrt{2}}\left(  \hat{G}_{1}-i\hat{G}_{2}\right),
\end{equation}
\begin{equation}
\hat{A}^{\dagger}=\frac{1}{\sqrt{2}}\left(  \hat{G}_{1}+i\hat{G}_{2}%
\right)  .
\end{equation}
These operators may also be obtained from the non Hermitian linear
invariant which arises from the complex solution of the TDHO
equation  \cite{Hacyan}. These annihilation and creation operators
are also invariant since they are composed by invariant operators.
On the other hand, the operators arising from (\ref{I gs}) yield
\begin{equation}
\hat{a}\left(  t\right)  =\frac{1}{\sqrt{2}}\left(
\frac{G\hat{q}}{\rho}+i(\rho\hat{p}-\dot{\rho}\hat{q})\right)
,\end{equation}
\begin{equation}
\hat{a}^{\dagger}\left(  t\right) =\frac{1}{\sqrt{2}}\left(
\frac{G\hat{q}}{\rho}-i(\rho\hat{p}-\dot{\rho}\hat
{q})\right)  .\label{ann}%
\end{equation}
These time dependent annihilation and creation operators were
originally introduced by Lewis  \cite{Lewis}. The Ermakov invariant
in terms of these operators is
\begin{equation}
\hat{I}=\hat{a}^{\dagger}\left(  t\right)  \hat{a}\left(  t\right)
+\frac
{G}{2}=\hat{A}^{\dagger}\hat{A}+\frac{G}{2},\label{a dagger a}%
\end{equation}
where the second equality follows from the definition of this
invariant in terms of the orthogonal functions quantum invariants
(\ref{I gs}). The time dependent annihilation (creation) operators
may be written as the product of the time independent annihilation
(creation) operators times a phase that only involves a
\textit{c}-number function. This expression may be written as a
unitary transformation of a phase shift
\begin{equation}
\hat{a}=\exp\left(  is_{\rho}\hat{I}\right)  \hat{A}\exp\left(
-is_{\rho
}\hat{I}\right)  \label{relat}%
\end{equation}
such that the equation of motion of this operator is then
\begin{equation}
\dot{\hat{a}}=i\omega(t)[\hat{I},\hat{a}].
\end{equation}

It is thus seen that the operator $ \omega(t)  \hat{I}$ in the
QM-TDHO plays the role that the Hamiltonian does in the time
independent harmonic oscillator case. This assertion is consistent
with two previous results. On the one hand, the transformation
that relates the invariant and the time dependent
Hamiltonian  \cite{gmc}
\begin{equation}
\omega(t)\hat{I}=\hat{H}(t)-i\frac{\partial
\hat{T}^{\dagger}}{\partial t}\hat{T},
\end{equation}
where the {\it squeeze} transformation   \cite{Vidiella} is given by
\begin{equation}
\hat{T}=\exp\left(
i\frac{\ln\rho\sqrt{\omega_0}}{2}(\hat{q}\hat{p}+\hat{p}\hat{q})\right)
\exp\left(  -i\frac{\dot{\rho}}{2\rho}\hat{q}^{2}\right)  .
\end{equation}
On the other hand, the propagator that describes the time
evolution of the transformed wave function
$|\psi(t)\rangle=\hat{U}_{I}\hat{T}^{\dagger}\hat
{T}(0)|\psi(0)\rangle$ in the time dependent case is
$\hat{U}_{I}=\exp\left( -is_{\rho}\hat{I}\right)  $. In the time
independent case of course $\hat{U}_{I}$ becomes the unity
operator. It is thus clear that the invariant in the time
dependent case enters the propagator expression in an analogous
fashion as the Hamiltonian does in the time independent case.

The energy of a time dependent classical oscillator in the
adiabatic approximation is proportional to
$\mathcal{E}\propto\rho^{2}\left(  t\right) \omega^{2}\left(
t\right)$. Therefore from (5) it is seen that the energy is
proportional to the frequency $\mathcal{E} \propto \omega(t)$
  \cite{fer2} whereas the Lewis constant has a quadratic of energy
over frequency dependence. However, the quantum versions of the
orthogonal functions invariants produce a linear form in the
coordinate and momentum operators as seen in Eqs. (\ref{gs}). It
has been necessary to evaluate the square of these operators in
order to obtain expressions proportional to the Hamiltonian of the
system. Therefore, a quantum invariant with a quadratic dependence
on the coordinate and momentum variables should be in
correspondence with the classical orthogonal functions invariant.

\section{Phase operator for time dependent coherent states}

Coherent states are states that follow classical trajectories and
are a standard to define non-classical features of quantum states.
Maybe because of their classicality, they have been useful to
describe phase in quantum optics  \cite{Lynch,Turski}. The invariant
formalism will be applied here to the phase operator given by
Turski. This operator will allow an appropriate translation of the
classical amplitude-phase invariant into the quantum one.

By using the annihilation operator (\ref{ann}) the
\textit{displacement} operator can be written as
$\hat{D}(\alpha)=\exp{(\alpha}\hat{a}{^{\dagger
}-\alpha^{\ast}}\hat{a}{)}$, $\alpha=r\exp(i\theta)$. The vacuum
state may
then be displaced to obtain a coherent state $|\alpha\rangle=\hat{D}%
(\alpha)|0\rangle$ and the phase operator introduced by
Turski  \cite{Turski}
is then generalized to the time dependent case%

\begin{equation}
\hat{\Phi}=\int\theta|\alpha\rangle\langle\alpha|d^{2}\alpha.
\end{equation}
This operator obeys the commutation relation
$[\hat{\Phi},\hat{I}]=-i$. In order to evaluate the time evolution
of $\hat{\Phi}$, this operator can be written in terms of the
invariant annihilation and
creation operators using (\ref{relat})%

\begin{equation}
\hat{\Phi}=e^{is_{\rho}\hat{I}}\left(
\int\theta\hat{D}_{A}(\alpha
)e^{-is_{\rho}\hat{I}}|0\rangle\langle0|e^{is_{\rho}\hat{I}}\hat{D}%
_{A}^{\dagger}(\alpha)d^{2}\alpha\right)  e^{-is_{\rho}\hat{I}},
\end{equation}
where $\hat{D}_{A}(\alpha)=\exp{(\alpha}\hat{A}{^{\dagger}-\alpha^{\ast}%
}\hat{A}{)}$. The invariant acting over the vacuum state is
$\hat{I}|0\rangle =\frac{1}{2}|0\rangle$ and the phase is then
\begin{equation}
\hat{\Phi}=e^{is_{\rho}\hat{I}}\left(
\int\theta\hat{D}_{A}(\alpha
)|0\rangle\langle0|\hat{D}_{A}^{\dagger}(\alpha)d^{2}\alpha\right)
e^{-is_{\rho}\hat{I}},
\end{equation}
the time derivative of this expression yields the equation of
motion for $\hat{\Phi}$:
\begin{equation}
\dot{\hat{\Phi}}=i\omega(t)[\hat{I},\hat{\Phi
}]=-\omega(t).\label{phievol}%
\end{equation}

The operator $\omega(t)\hat{I}$ once again takes the role of the
Hamiltonian.

\section{Number and phase operators}

The coordinate operator from (\ref{ann}) is
\begin{equation}
\hat{q}=\sqrt{\frac{1}{2G\omega(t)}}(\hat{a}+\hat
{a}^{\dagger}), \label{qq}%
\end{equation}
and following Dirac  \cite{dirac} the creation and annihilation
operators may be written as
\begin{equation}
\hat{a}=\sqrt{\hat{I}}e^{-i\hat{\Phi}},\qquad\hat{a}^{\dagger}=e^{i\hat{\Phi}%
}\sqrt{\hat{I}}. \label{dira}%
\end{equation}
The coordinate operator (\ref{qq}) in the form of amplitude and
phase variables is then
\begin{equation}
\hat{q}=\sqrt{\frac{\hat{I}}{2G\omega(t)}}e^{-i\hat{\Phi
}}+e^{i\hat{\Phi}}\sqrt{\frac{\hat{I}}{2G\omega(t)}},
\label{qqq}%
\end{equation}
where the amplitude $\rho$ and phase $s_{\rho}$ are identified as
\begin{equation}
\rho\rightarrow\sqrt{\frac{\hat{I}}{G\omega(t)}},\qquad
s_{\rho}\rightarrow\hat{\Phi}.
\end{equation}

The invariant with the aid of (\ref{phievol}) is given in
amplitude and phase operators as
\begin{equation}
\hat{I}=-\frac{\hat{a}^{\dagger}\hat{a}+\frac{1}{2}%
}{G\omega(t)}\dot{\hat{\Phi}}, \label{quant inv}%
\end{equation}
which has the same structure of the orthogonal functions classical
invariant written in amplitude and phase variables (5). The number operator is then identified with
\begin{equation}
\hat{n}\left(  t\right) =\frac{1}{G\omega(t)}\hat
{a}^{\dagger}\hat{a}. \label{numb op}%
\end{equation}

It should be recalled that the number operator is identified with
the product of the annihilation and creation pair when these
operators arise from the Hamiltonian operator. However, the
identification of the number operator in terms of the annihilation
- creation pair arising from the invariant is obtained from the
correspondence with the orthogonal functions classical invariant.
The invariant in terms of the number and phase operators is then
\begin{equation}
\hat{I}=-\left(  \hat{n}\left(  t\right) +\frac{1}{2}\frac
{1}{\omega(t)}\right) \dot{\hat{\Phi}}\left(  t\right)
.\label{inv n phi dot}%
\end{equation}
Since $\hat{a}^{\dagger}\hat{a}$ is invariant from (\ref{a dagger
a}), if the frequency is constant the number of excitations is
then also constant. Nonetheless, in the time dependent case, the
number of excitations is inversely proportional to the time
dependent frequency in correspondence with the intensity
dependence obtained in the classical limit.

The energy of the excitation at a given time $t_{s}$ is given by
$\mathcal{E}=\hat{n}\left(  t_{s}\right)  \omega\left(
t_{s}\right)  $ (with $\hbar=1$) but this is precisely the quantum
invariant value above the vacuum state $\hat{I}=-$ $\hat{n}\left(
t_{s}\right)  \dot{\hat{\Phi}}\left( t_{s}\right)  $. Therefore,
the invariant represents the energy conservation of the closed
system. In contrast, the time dependent Hamiltonian is no longer a
constant of motion whose eigenvalue is necessarily related to an
open system.

\section{Conservation equations}

Consider, as an example of this formalism, the number of
excitations to represent the photon number. Let an ensemble of
oscillators be followed along their propagating path so that the
properties of the oscillators are time dependent quantities. This
scheme corresponds to a Lagrangian hydrodynamic framework.

Allow for a nonlinear process where the wave mixing at an initial
time consists of $\hat{n}_{1}\left(  t_{i}\right)
,\hat{n}_{2}\left( t_{i}\right)  ,...,\hat{n}_{n}\left(
t_{i}\right)  $ oscillators with frequencies $\omega_{1}\left(
t_{i}\right)  ,\omega_{2}\left(  t_{i}\right)
,...,\omega_{n}\left(  t_{i}\right)  $. The system then evolves to
frequencies $\omega_{1}\left(  t_{f}\right)  ,\omega_{2}\left(
t_{f}\right) ,...,\omega_{n}\left(  t_{f}\right)  $ with
$\hat{n}_{1}\left(  t_{f}\right) ,\hat{n}_{2}\left(  t_{f}\right)
,...,\hat{n}_{n}\left(  t_{f}\right)  $ photons at a time $t_{f}$.
For any  pair it is possible to establish the corresponding
invariant
relationship (\ref{inv n phi dot}), thus%
\begin{equation}
\omega_{k}\left(  t_{i}\right)  \hat{n}_{k}\left(  t_{i}\right)
=\omega _{k}\left(  t_{f}\right)  \hat{n}_{k}\left(  t_{f}\right)
\quad k\in1,2,...,n.
\end{equation}
Each of these equations is stating that the energy of an ensemble
of $\hat {n}_{k}\left(  t_{i}\right)  $ oscillators with frequency
$\omega_{k}\left( t_{i}\right)  $ is equal to the energy of
$\hat{n}_{k}\left(  t_{f}\right)  $ oscillators with frequency
$\omega_{k}\left(  t_{f}\right)  $.

The particular nonlinear process taking place then imposes a
relationship between the input and output frequencies. For
example, sum frequency generation used to produce VUV tunable
radiation  \cite{Hanna} implies that
\begin{equation}
2\omega_{1}+\omega_{2}=\omega_{f}.
\end{equation}
In the above notation this expression corresponds to%
\begin{equation}
2\omega_{1}\left(  t_{i}\right)  +\omega_{2}\left(  t_{i}\right)
=\omega _{1}\left(  t_{f}\right)  =\omega_{2}\left(  t_{f}\right)
.
\end{equation}
\ If we multiply this equation by the number of photons in the
final state
$\hat{n}_{1}\left(  t_{f}\right)  $, the equation reads%
\begin{equation}
2\omega_{1}\left(  t_{i}\right)  \hat{n}_{1}\left(  t_{f}\right)
+\omega _{2}\left(  t_{i}\right)  \hat{n}_{1}\left(  t_{f}\right)
=\omega_{1}\left( t_{f}\right)  \hat{n}_{1}\left(  t_{f}\right)  ,
\end{equation}
but since $\omega_{1}\left(  t_{i}\right)  \hat{n}_{1}\left(
t_{i}\right) =\omega_{1}\left(  t_{f}\right)  \hat{n}_{1}\left(
t_{f}\right)  $ and $\omega_{2}\left(  t_{i}\right)
\hat{n}_{2}\left(  t_{i}\right)  =\omega _{2}\left(  t_{f}\right)
\hat{n}_{2}\left(  t_{f}\right)  $, the above
expression may be casted as%
\begin{equation}
2\omega_{1}^{2}\left(  t_{i}\right)  \frac{\hat{n}_{1}\left(
t_{i}\right) }{\omega_{1}\left(  t_{f}\right)
}+\omega_{2}^{2}\left(  t_{i}\right) \frac{\hat{n}_{2}\left(
t_{i}\right)  }{\omega_{1}\left(  t_{f}\right) }=\hat{n}_{1}\left(
t_{f}\right)  \omega_{1}\left(  t_{f}\right)  .
\end{equation}
This result is often cited in the literature in terms of the power
density, defined as the energy per unit time
$\mathcal{W}_{k}=\mathcal{E}_{k}/\tau
_{k}=\omega_{k}^{2}\hat{n}_{k}$, at these two times is then given
by
\begin{equation}
\mathcal{W}_{1}\left(  t_{i}\right)  +\mathcal{W}_{2}\left(
t_{i}\right) =\mathcal{W}_{1}\left(  t_{f}\right)  .
\end{equation}

This reasoning may be applied to any other nonlinear processes
such as parametric amplification or frequency difference.

\section{Conclusions}

The Ermakov Lewis invariant may be used in an equivalent fashion
as the Hamiltonian is used in the time independent case. Namely,
to obtain evolution operators, to cast the equations of motion of
different operators in commutative expressions, and to produce a
phase shift with its exponential form. The number and phase
representation of this invariant corresponds to the classical
orthogonal functions invariant, which in turn is proportional to
the ratio of energy over frequency in the adiabatic approximation.
These properties of the Ermakov Lewis invariant suggest that this
quantity is proportional to the total energy of the closed system.

A quantum derivation of the conservation equations has been
obtained for arbitrary non-linear frequency conversion processes.
These results in a Lagrangian frame of reference are analogous to
the flux conservation in the semiclassical Eulerian framework. An
important difference between the two descriptions is that in the
semiclassical case, the fields at different fixed frequencies
change their amplitudes as a function of position. In contrast,
this quantum treatment considers the time evolution of the
frequency with its corresponding amplitude variation.


\begin{thebibliography}{99}
\bibitem{Ray}J.R. Ray, Phys. Rev. A \textbf{26} (2) 729 (1982).

\bibitem{Lutzky}M. Lutzky, Phys. Lett. \textbf{68A} 3 (1978).

\bibitem{gmc} M. Fern\'{a}ndez Guasti and H. Moya-Cessa, J. Phys. A \textbf{36}
2069 (2003); H Moya-Cessa, M Fern\'andez Guasti, Phys. Lett. A {\bf 311}  1 (2003).

\bibitem{fermoy pra}M. Fern\'{a}ndez Guasti and H. Moya-Cessa, Phys. Rev. A
\textbf{67} 063803 (2003). 

\bibitem{Eliezer}C.J. Eliezer and A. Gray, SIAM J. Appl. Math. \textbf{30} (3)
463 (1976).

\bibitem{fer2}M. Fern\'{a}ndez Guasti and A. Gil-Villegas, in \textit{Recent
Developments in Mathematical and Experimental Physics}, edited by
A. Macias, F. Uribe and E. Diaz, (Vol. C: Hydrodynamics and
Dynamical Systems, Kluwer, NY, 2003) pp.159-166.

\bibitem{Manley}J.M. Manley, H.E. Rowe, Proc. IRE \textbf{47}, 2115 (1959)

\bibitem{Yariv}A. Yariv, \textit{Quantum electronics}, 2nd ed. (J. Wiley, NY,
1989), p.410

\bibitem{Shen}Y.R. Shen, \textit{The principles of Nonlinear Optics}, (J.
Wiley, NY, 1984), p.78

\bibitem{Planck}T.S. Kuhn, \textit{Black-Body Theory and the Quantum
Discontinuity 1894-1912}, (Chicago, University of Chicago Press,
1987)

\bibitem{fer1}M. Fern\'{a}ndez Guasti and A. Gil-Villegas, Phys. Lett. A
\textbf{292}, (4-5), 243 (2002).

\bibitem{Lewis}H.R. Lewis, Phys. Rev. Lett. \textbf{18} 510 (1967).

\bibitem{Hacyan}S. Hacyan and R. Jauregui,  J. Opt B \textbf{5} 138
(2003).

\bibitem{Vidiella} H. Moya-Cessa and A. Vidiella-Barranco J. of Mod. Optics
{\bf 42}, 1547 (1995).

\bibitem{Lynch}R. Lynch, Phys. Rep. \textbf{256} 367 (1995)

\bibitem{Turski}L.A Turski, Physica \textbf{57} 432 (1972); see also M.L.
Arroyo Carrasco and H. Moya-Cessa, Quant. Semiclass. Opt.
\textbf{9}, L1 (1997).

\bibitem{Hanna}D.C. Hanna, M.A. Yuratrich, D. Cotter \textit{Nonlinear Optics
of Free Atoms and Molecules} (Berlin, Springer Verlag, 1979) p.
137

\bibitem{Ray and Reid}J.R. Ray and J.L. Reid, Phys. Rev. A \textbf{26} (2)
1042 (1982).

\bibitem{vogel}W. Vogel and D.-G. Welsch \textit{Lectures on Quantum Optics},
(Berlin, Akad. Verl., 1994).

\bibitem{dirac}P.A.M. Dirac, Proc. R. Soc. A \textbf{114}, 243 (1927).
\end{thebibliography}
\end{document}